\documentclass[a4paper,12pt]{article}
\usepackage{amsmath}
\usepackage{amsfonts,amssymb}
\usepackage{preprint}
\begin{document}
\renewcommand{\theequation}{\arabic{equation}}
\baselineskip=14pt  
\abovedisplayskip=10pt
\belowdisplayskip=10pt
\jot=5pt
%
%
\def\calO{{\cal O}}
\def\boldN{{\mathbf{N}}}
\def\boldC{{\mathbf{C}}}
\def\bolda{\hbox{\boldmath$\mathit{a}$}}
\def\calI{{\cal I}}
%
\hrule height0pt depth0pt
\vspace*{-12pt}
\rightline{\bf RIMS-1334}
\vspace*{50pt}
\centerline{\LARGE
Nonlinear Transformation Group of CAR Fermion Algebra}
\vskip100pt
\renewcommand{\thefootnote}{\alph{footnote})}
\centerline{\large
 Mitsuo Abe\footnote{E-mail: \abemail}
 and Katsunori Kawamura\footnote{E-mail: \kkmail}
}
\vskip10pt
\centerline{\it Research Institute for Mathematical Sciences,
Kyoto University, Kyoto 606-8502, Japan}
\vskip25pt
\vskip120pt
\centerline{\itshape\bfseries Abstract}
\vskip5pt
Based on our previous work on the recursive fermion system in the Cuntz 
algebra, it is shown that a nonlinear transformation group of the CAR fermion 
algebra is induced from a $U(2^p)$ action on the Cuntz algebra $\calO_{2^p}$ 
with an arbitrary positive integer $p$. 
In general, these nonlinear transformations are expressed in terms of
finite polynomials in generators.
Some Bogoliubov transformations are involved as special cases.
\vfill\eject
%
%
%
\par
In our previous paper,\cite{AK1} we have introduced the recursive fermion 
system (RFS$_p$) which gives embeddings of CAR into the Cuntz 
algebra\cite{Cuntz} $\calO_{2^p}$ with an arbitrary positive integer $p$. 
As for the special case, which we call the {\it standard\/} RFS$_p$, 
CAR is embedded onto the $U(1)$-invariant $\ast$-subalgebra 
$\calO_{2^p}^{U(1)}$ of $\calO_{2^p}$.  Here, the $U(1)$ action on $\calO_d$ 
is defined by 
\begin{equation}
\gamma_z : s_i \mapsto z\,s_i, \quad z\in\boldC,\quad  |z|=1
\end{equation}
with $s_i$ $(i=1,\,\ldots,\,d)$ being the generators of  $\calO_d$. 
Since an automorphism of $\calO_{2^p}$, which is described by a $U(2^p)$ 
action, also keeps $\calO_{2^p}^{U(1)}$ invariant, it induces a 
transformation of CAR. 
We find this type of transformations is {\it nonlinear\/} with respect to 
generators $\{a_m,\,a_n^* \, | \,  m,\,n = 1,\,2,\,\ldots\,\}$ in general, 
and includes some Bogoliubov transformations as special cases.
\par
First, let us recall that the Cuntz algebra $\calO_d$ is a $C^*$-algebra 
generated by $s_i, \, i=1,\,2,\,\ldots,\,d$, which satisfy the following 
relations:\cite{Cuntz}
\begin{eqnarray}
&& s_i^*\, s_j = \delta_{i,j}I, \\
&& \sum_{i=1}^{d} s_i\, s_i^* =I.
\end{eqnarray}
We often use the brief description such as 
$s_{i_1i_2\cdots i_m;\,j_n\,\cdots\, j_2\,j_1}
\equiv s_{i_1} s_{i_2} \cdots s_{i_m} s_{j_n}^* \cdots s_{j_2}^* s_{j_1}^*$,
$m+n\geqq1$.
The $U(1)$-invariant subalgebra $\calO_d^{U(1)}$ of $\calO_d$ is given by 
a linear space spanned by $s_{i_1\cdots i_m;\,j_m\cdots j_1}$, 
$m=1,\,2,\,\ldots\,$.
We consider an automorphism $\alpha_u$ of $\calO_d$ obtained from a natural 
$U(d)$ action $\alpha$ as follows:
\begin{eqnarray}
&&
\begin{array}{c}
\alpha : U(d) \curvearrowright \, \calO_d, \\[5pt]
\alpha_u (s_i) \equiv \sum\limits_{k=1}^{d} s_{k} u_{k i}, \quad 
     u \in U(d), \quad i=1,\,2,\,\ldots,\,d.  
\end{array}
\end{eqnarray}
Indeed, from the equality
\begin{eqnarray}
&&\alpha_{u_1}\circ\alpha_{u_2}=\alpha_{u_1u_2}, \quad 
     u_1, \, u_2 \in U(d),
\end{eqnarray}
$\alpha$ becomes a $U(d)$ action on $\calO_d$.
Since $\alpha_u\circ\gamma_z=\gamma_z\circ\alpha_u$, 
restriction of $\alpha_u$ to $\calO_d^{U(1)}$ gives an automorphism 
of $\calO_d^{U(1)}$:
\begin{eqnarray}
&&
\begin{array}{c}
\alpha_u|_{O_d^{U(1)}} \in \mbox{Aut}\, \calO_d^{U(1)}, \\[5pt]
\alpha_u(s_{i_1\cdots i_m;\,j_m\cdots j_1})
=\sum\limits_{k_1,\ldots,\ell_m=1}^{d} 
   u_{k_1i_1}\cdots u_{k_m i_m} u^*_{j_1\ell_1}\cdots u^*_{j_m \ell_m}
   s_{k_1\cdots k_m;\, \ell_m\cdots \ell_1} \in O_d^{U(1)}.
\end{array}                                             \label{aut-Od}
\end{eqnarray}
Using the homogeneous embedding $\varPsi$ of $\calO_{d^r}$ 
$(r=2,\,3,\,\ldots)$ into $\calO_d$ defined by\cite{AK2}
\begin{eqnarray}
&&
\begin{array}{c}
\varPsi : \calO_{d^r} \hookrightarrow \calO_d, \\[8pt]
\varPsi(S_i) \equiv s_{j_1}\cdots s_{j_r}, \quad i=1,\,2,\,\ldots\,,d^r; \ 
   j_1, \, \ldots\,, j_r = 1,\,2,\,\ldots\,,d, \\[2pt]
\quad 
   i - 1 = \sum\limits_{k=1}^r (j_k - 1) d^{k-1},
\end{array}
\end{eqnarray} 
where $S_1, \, \ldots\,, S_{d^r}$ and $s_1, \, \ldots\,, s_d$ denote
the generators of $\calO_{d^p}$ and $\calO_d$, respectively, 
it is straightforward to show that, for  any $u\in U(d)$, there exists 
$v\in U(d^r)$ such that
\begin{equation}
(\alpha_u\circ\varPsi)(X)=(\varPsi\circ\alpha_v)(X), 
\quad X \in \calO_{d^r}.                                 \label{unitary-embed}
\end{equation}
\par
The generators of CAR are denoted by $a_n$, $n=1,\,2,\,\dots\,$, which 
satisfy
\begin{equation}
\{a_m, \, a_n\}=0, \quad \{ a_m, \, a_n^* \} = \delta_{m,n} I, \quad
m,\, n = 1,\,2,\,\ldots.
\end{equation}
We give a systematic embedding of CAR into $\calO_{2^p}$ in the following.
The {\it standard recursive fermion system of order $p$\/} (RFS$_p$) in 
$\calO_{2^p}$ is given by a set 
$SR_p=(\bolda_1, \,\ldots, \, \bolda_p\,;\, \zeta_p, \, \varphi_p )$ 
defined by\cite{AK1}
\begin{eqnarray}
&& \bolda_i \equiv \sum_{k=1}^{2^{p-i}}\sum_{\ell=1}^{2^{i-1}}
   (-1)^{\sum\limits_{m=1}^{i-1}\left[\frac{\ell-1}{2^{m-1}}\right]}
   s_{2^i(k-1)+\ell}s_{2^{i-1}(2k-1)+\ell}^*, 
   \quad i=1,\, 2,\, \ldots, \, p,                      \label{RFSp-1}\\
&& \zeta_p(X) \equiv \sum_{i=1}^{2^p}
   (-1)^{\sum\limits_{m=1}^{p}\left[\frac{i-1}{2^{m-1}}\right]}
   s_i X s_i^*,  \quad X \in \calO_{2^p},               \label{RFSp-2}\\
&& \varphi_p(X) \equiv \rho_{2^p}(X) \equiv \sum_{i=1}^{2^p} s_i X s_i^*,
   \quad X \in \calO_{2^p},
\end{eqnarray}
where $s_i$'s $(i=1,\,2,\,\ldots,\,2^p)$ are generators of $\calO_{2^p}$,
$[x]$ denotes the largest integer not greater than $x$, and
$\rho_{2^p}$ is the canonical endomorphism of $\calO_{2^p}$. 
Here, $SR_p$ satisfies the following relations
\begin{eqnarray}
&& \{ \bolda_i, \, \bolda_j \} = 0, \quad  
   \{ \bolda_i, \, \bolda_j^* \} = \delta_{i,j}I, \quad
   i, \, j=1,\, 2,\, \ldots,\, p, \\[2pt]
&& \{ \bolda_i, \, \zeta_p(X) \} = 0, \quad
   \zeta_p(X)^* = \zeta(X^*), \quad 
   i=1,\, 2,\, \ldots, \, p, \ \ X \in \calO_{2^p},\\[2pt]
&& \zeta_p(X) \zeta_p(Y) = \varphi_p(XY), \quad 
   X, \, Y \in \calO_{2^p}.
\end{eqnarray}
Then, the embedding $\varPhi_{SR_p}$ of CAR into $\calO_{2^p}$ associated 
with $SR_p$ is defined by
\begin{equation}
\begin{array}{c}
\varPhi_{SR_p} : \mbox{CAR} \hookrightarrow \calO_{2^p},    \\[7pt]
\varPhi_{SR_p}( a_{p(n-1)+i} ) = \zeta_p^{n-1}( \bolda_i ), \quad
 i=1,\, 2,\,\ldots,\, p, \ \ n=1,\,2,\,\ldots. \\[5pt]
\end{array}
\end{equation}
For better understanding, we write explicitly $\bolda_i$ and $\zeta_p$ of 
$SR_p$ for $p=1, \, 2, \, 3$:
\begin{eqnarray}
&&
SR_1\, \left\{\begin{array}{rl}
\bolda_1 \equiv&\!\!\! s_{1;\,2}, \\[7pt]
\zeta_1(X) \equiv&\!\!\! s_1 X s_1^* - s_2 X  s_2^*,
  \quad X \in \calO_2\,;
\end{array}\right.\\[10pt]
&&
SR_2\, \left\{\begin{array}{rl}
\bolda_1 \equiv&\!\!\! s_{1;\,2} + s_{3;\,4}, \\[7pt]
\bolda_2 \equiv&\!\!\! s_{1;\,3} - s_{2;\,4}, \\[7pt]
\zeta_2(X)\equiv&\!\!\! s_1 X s_1^*-s_2 X s_2^*-s_3 X s_3^*+s_4 X s_4^*, 
          \quad X \in \calO_4\,;
\end{array}\right. \\ [10pt]
\noalign{\pagebreak}
&& 
SR_3\, \left\{\begin{array}{rl}
\bolda_1 \equiv&\!\!\! s_{1;\,2} + s_{3;\,4} + s_{5;\,6} + s_{7;\,8}, \\[7pt]
\bolda_2 \equiv&\!\!\! s_{1;\,3} - s_{2;\,4} + s_{5;\,7} - s_{6;\,8}, \\[7pt]
\bolda_3 \equiv&\!\!\! s_{1;\,5} - s_{2;\,6} - s_{3;\,7} + s_{4;\,8}, \\[7pt]
\zeta_3(X) \equiv&\!\!\! 
   s_1 X s_1^* - s_2 X s_2^* - s_3 X  s_3^* + s_4 X s_4^* \\[5pt]
   &\!\!\! -s_5 X s_5^* + s_6 X s_6^* + s_7 X s_7^* - s_8 X s_8^*,
       \quad X \in \calO_{8}.
\end{array}\right.
\end{eqnarray}
Using mathematical induction, it is shown that $\varPhi_{SR_p}(\mbox{CAR})$
is identical with $\calO_{2^p}^{U(1)}$, that is, any 
$s_{i_1\cdots i_m; \, j_m\,\cdots\, j_1}\in\calO_{2^p}^{U(1)}$ is 
expressed in terms of a monomial in $\varPhi_{SR_p}(a_k)$ and 
$\varPhi_{SR_p}(a_\ell^*)$ with $k,\ell\leqq p m$.
\par
From \eqref{aut-Od} with $d=2^p$ and using the fact 
$\calO_{2^p}^{U(1)}=\varPhi_{SR_p}(\mbox{CAR})$,
we have
\begin{equation}
\alpha_u|_{\varPhi_{SR_p}(\mbox{\scriptsize CAR})} 
 \in \mbox{Aut}\, \varPhi_{SR_p}(\mbox{CAR}), \quad u\in U(2^p).
\end{equation}
Hence, the automorphism $\alpha_u$ of $\calO_{2^p}$ induces 
a $U(2^p)$ action $\chi$ on CAR as follows
\begin{equation}
\begin{array}{c}
\chi : U(2^p) \curvearrowright \, \mbox{CAR}, \\[7pt]
\chi_u(a_n)\equiv
  (\varPhi_{SR_p}^{-1}\circ \alpha_u\circ\varPhi_{SR_p})(a_n),\quad 
  u\in U(2^p), \ \ n=1,\,2,\,\ldots\,. \\[7pt]
\end{array}                                             \label{nlt-car}
\end{equation}
Here, the rhs of \eqref{nlt-car} is expressed in terms of a finite polynomial 
in $a_k$ and $a_\ell^*$ with $k, \ell\leqq pm$ for $p(m-1)+1 \leqq n \leqq pm$.
Therefore, $\chi_u$ gives a nonlinear transformation of CAR associated with
$u\in U(2^p)$.
It is obvious that the whole set of $\chi_u$ with $u\in U(2^p)$ denoted by
\begin{eqnarray}
&&
\mbox{Aut}_{\,U(2^p)}(\mbox{CAR}) \equiv \{ \chi_u \, | \, u\in U(2^p) \}
\end{eqnarray}
constitutes a kind of nonlinear realization of $U(2^p)$ on CAR.
Restricting \eqref{unitary-embed} with $d=2^p$ to 
$\varPhi_{SR_{rp}}(\mbox{CAR})$, we obtain the following inclusion relation
\begin{eqnarray}
&&
\mbox{Aut}_{\,U(2^p)}(\mbox{CAR}) \subset 
\mbox{Aut}_{\,U(2^{rp})}(\mbox{CAR}), \quad 
  r=2,\,3,\,\ldots.                                     \label{inclusion}
\end{eqnarray}
Hence, it may be possible to define some kind of inductive limit as
Aut$_{\,U(2^\infty)}(\mbox{CAR})$, but we will not proceed in detail here. 
\par
Although the $U(2^p$) action $\chi$ on CAR is defined individually for 
each $p\in\boldN$, it is possible to consider a product 
$\chi_{u_1}\circ\chi_{u_2}$ even for $u_1\in U(2^{p_1})$ and 
$u_2\in U(2^{p_2})$ with $p_1\not=p_2$ according to its definition.  
From \eqref{inclusion}, this product is also given explicitly in 
Aut$_{U(2^q)}(\mbox{CAR})$ with $q$ being a common multiple of $p_1$ and 
$p_2$.  Indeed,  there exist $v_1, \, v_2 \in U(2^q)$ satisfying 
$\chi_{u_i}=\chi_{v_i}$, $i=1,\,2$, hence, we have 
$\chi_{u_1}\circ\chi_{u_2}=\chi_{v_1 v_2}$. 
In any way, we obtain an infinite-dimensional automorphism group of CAR 
by giving the whole set of such products as follows
\begin{equation}
\mbox{Aut}_U(\mbox{CAR})\equiv\{ \, 
 \chi_{u_1}\circ\cdots\circ\chi_{u_k}\  | \  
 u_i\in U(2^{p_i}),\ i=1,\,2,\,\ldots,\,k\,;\ p_i,\,k=1,\,2,\,\ldots\,\}.
\end{equation}
\par
Hereafter, we consider the nonlinear transformation of CAR defined
by \eqref{nlt-car} more concretely.  We rewrite \eqref{RFSp-2} as  
$\zeta_p(X)=\sum\limits_{i\in\calI_+} s_i X s_i^* 
- \sum\limits_{i\in\calI_-} s_i X s_i^*$ with $\calI_\pm$ being a certain
division of indices $\{ 1, \,2,\,\ldots,\,2^p \}$ satisfying 
$\sharp\,\calI_\pm=2^{p-1}$.
Then, for any $u\in U(2^p)\bigcap U(2^{p-1}, 2^{p-1})
=U(2^{p-1})\times U(2^{p-1})$, 
where each $U(2^{p-1})$ acts only on $s_i$ with $i\in \calI_+$ 
(or $i\in\calI_-$), 
we have an equality $\alpha_u(\zeta_p(X))=\zeta_p(\alpha_u(X))$, that is, 
$\zeta_p$ is {\it covariant\/} with respect to $\alpha_u$ for 
$u\in U(2^{p-1})\times U(2^{p-1})$.
On the other hand, if $k$ is an odd positive integer, we have an identity 
$\zeta_p(X_1\cdots X_k)=\zeta_p(X_1)\cdots\zeta_p(X_k)$ 
$(X_1,\,\ldots,\,X_{k} \in \calO_{2^p})$.
Therefore, if $u\in U(2^{p-1})\times U(2^{p-1})$ and $\alpha_u(\bolda_i)$ 
involves only odd order monomials in $\bolda_i$ $(i=1,\,\ldots,\,p)$, 
$\{\alpha_u(\varPhi_{SR_p}(a_{p(n-1)+i})) \,|\, \ i=1,\,2,\,\ldots,\,p\}$ 
for each $n$ is expressed in terms of polynomials in the same form as 
$\{\alpha_u(\bolda_i) \,|\, i=1,\,2,\,\ldots,\,p\}$, that is, $\varPhi_{SR_p}$ 
is {\it covariant\/} with respect to $\alpha_u$.
\par
For example, for $p\geqq2$, let $u$ be the transposition of $s_{2^{p-1}+1}$ 
and $s_{2^p}$. Then, from an identity $\bolda_1 \bolda_2 \cdots \bolda_p 
= (-1)^{\left[\frac{p}{2}\right]}s_{1, 2^p}$,
$\alpha_u(\bolda_p)$ is expressed in terms of a polynomial involving
a term $(-1)^{\left[\frac{p}{2}\right]}\bolda_1\cdots\bolda_p$. 
Furthermore, if $p$ is odd, not only $\zeta_p$ but also $\varPhi_{SR_p}$ 
is covariant with respect to this $\alpha_u$.
\par
In the following, we present some examples explicitly. 
\begin{list}{}{\topsep=10pt \itemsep=10pt \parsep=0pt}
\item[(1)]
$p=1$, $u=u_1(\theta) \in SO(2)\subset U(2)$;
$\alpha_{u_1(\theta)}(s_1)\equiv \cos\theta\,s_1 - \sin\theta\,s_2$,
$\alpha_{u_1(\theta)}(s_2)\equiv \sin\theta\,s_1 + \cos\theta\,s_2$, 
 $(0\leqq\theta<2\pi)$:
\begin{eqnarray}
\chi_{u_1(\theta)}(a_1)
&\!=\!&   \cos^2 \theta\, a_1 -  \sin^2 \theta\, a_1^* 
    + \sin\theta \cos\theta\,(I-2a_1^*a_1) \nonumber \\
&\!=\!&   \cos^2\theta\, a_1- \sin^2\theta\, a_1^* 
    + \sin\theta\cos\theta\exp(i\pi a_1^* a_1), \\
\chi_{u_1(\theta)}(a_n)
&\!=\!&\prod_{k=1}^{n-1}\Big( 
      \cos2\theta 
    - \sin2\theta\,\exp\!\Big(i\pi\sum_{i=1}^{k-1} a_i^* a_i\Big)
       (a_k^* - a_k) \Big) \nonumber\\
&&\! \quad\times\!\left( 
      \cos^2\theta\, a_n- \sin^2\theta\, a_n^* 
 + \sin\theta\cos\theta\exp\!\Big(i\pi \sum_{j=1}^n a_j^* a_j\Big)\!\right)\!,
      \  n\geqq2,\qquad
\end{eqnarray}
where use has been made of an identity $(a_k^* a_k)^2 = a_k^* a_k$, and 
$\exp(i\pi a_k^* a_k)$ being the Klein operator anticommuting with $a_k$.
One should note that an identity $\exp(i2\pi a_k^* a_k)=I$ holds.
From \eqref{inclusion}, the same nonlinear transformation is involved
in Aut$_{\,U(2^p)}(\mbox{CAR})$ for any $p\in\boldN$. 
Indeed, it is straightforward to show it for $p=2$.
\item[(2)]
$p=2$, $u=u_2\in\mathfrak{S}_4\subset U(4)$; 
$\alpha_{u_2}(s_3)\equiv s_4$,  
$\alpha_{u_2}(s_4)\equiv s_3$, 
$\alpha_{u_2}(s_i)\equiv s_i$ for $i=1,\,2$: 
\begin{alignat}{2}
\chi_{u_2}(a_1)
 &= a_1 + ( a_1^* - a_1 ) a_2^* a_2, &\\
\chi_{u_2}(a_2)
 &= - (a_1^* + a_1 ) a_2,  &\\
\chi_{u_2}( a_{2n-1} )
&= \phantom{-} \exp\!\Big(i\pi\sum_{k=1}^{n-1} a_{2k}^* a_{2k}\Big) 
   \Big( a_{2n-1} + ( a_{2n-1}^* - a_{2n-1} ) a_{2n}^* a_{2n} \Big),  
    &\ \ n\geqq2,\\
\chi_{u_2}(a_{2n}) 
 &= - \exp\!\Big(i\pi\sum_{k=1}^{n-1} a_{2k-1}^* a_{2k-1}\Big)
             ( a_{2n-1}^* + a_{2n-1} ) a_{2n},    &\ \ n\geqq2.
\end{alignat}
It is straightforward to confirm that the equality 
$(\alpha_{u_2}\circ\alpha_{u_2})(a_n)=a_n$ is satisfied, 
which is the result of $u_2=u_2^{-1}$.  
\item[(3)]
$p=2$, $u=u_3\in\mathfrak{S}_4\subset U(4)$; 
$\alpha_{u_3}(s_2)\equiv s_3$, 
$\alpha_{u_2}(s_3)\equiv s_2$, 
$\alpha_{u_3}(s_i)\equiv s_i$ for $i=1,\,4$: 
\begin{eqnarray}
\chi_{u_3}( a_{2n-1} )
 &=& \exp\!\Big(i\pi a_{2n-1}^* a_{2n-1}\Big) a_{2n}, \\
\chi_{u_3}(a_{2n}) 
 &=& \exp\!\Big(i\pi a_{2n}^* a_{2n}\Big) a_{2n-1}.
\end{eqnarray}
This transformation should be called a double Klein transformation which
preserves anticommutativity between two annihilation (creation) operators.
\item[(4)]
$p=2$, $u=u_4(\theta) \in SO(2)\subset U(4) $;
$\alpha_{u_4(\theta)}(s_1)\equiv\cos\theta s_1 - \sin\theta s_4$, 
$\alpha_{u_4(\theta)}(s_4)\equiv\sin\theta s_1 + \cos\theta s_4$, 
 $(0\leqq\theta<2\pi)$,
$\alpha_{u_4(\theta)}(s_i)=s_i$ for $i=2,\,3$: 
\begin{eqnarray}
\chi_{u_4(\theta)}(a_{2n-1}) 
  &=& \phantom{-} \cos\theta\, a_{2n-1} + \sin\theta \, a_{2n}^*, \\
\chi_{u_4(\theta)}(a_{2n})    
  &=& -\sin\theta\, a_{2n-1}^* + \cos\theta \, a_{2n}.
\end{eqnarray}
This is nothing but an example of the conventional Bogoliubov transformation.
Similar Bogoliubov transformations are obtained in 
Aut$_{\,U(4^r)}(\mbox{CAR})$ with $r\in\boldN$.
\item[(5)]
$p=3$, $u=u_5 \in \mathfrak{S}_8\subset U(8)$;
$\alpha_{u_5}(s_5)\equiv s_8$, 
$\alpha_{u_5}(s_8)\equiv s_5$, 
$\alpha_{u_5}(s_i)\equiv s_i$ for $i\not=5,\,8$: 
\begin{eqnarray}
\chi_{u_5}(a_{3n-2})
 &=& a_{3n-2} - a_{3n-2} a_{3n}^* a_{3n} 
        + ( I - 2 a_{3n-2}^* a_{3n-2}) a_{3n-1}^* a_{3n}^* a_{3n}, \\
\chi_{u_5}(a_{3n-1})
 &=& a_{3n-1} - a_{3n-2}a_{3n}^* a_{3n} 
         - a_{3n-2}^* ( I - 2 a_{3n-1}^* a_{3n-1}) a_{3n}^* a_{3n}, \\
\chi_{u_5}(a_{3n})
 &=& -(a_{3n-2}a_{3n-1} + a_{3n-2}^* a_{3n-2} + a_{3n-1}^* a_{3n-1}\nonumber\\
 && \hspace*{50pt} 
   + a_{3n-2}^* a_{3n-1}^* - 2 a_{3n-2}^* a_{3n-2} a_{3n-1}^* a_{3n-1})a_{3n}.
\end{eqnarray}
This is an example mentioned above for the case $p=3$. 
Here, $\chi_{u_5}(a_{3n})$ involves a term 
$-a_{3n-2}a_{3n-1}a_{3n}$, and $\varPhi_{SR_3}$ is covariant with 
respect to $\alpha_{u_5}$.
\end{list}
\par
In conclusion, we summarize the properties of the above nonlinear 
transformations constituting $Aut_U(\mbox{CAR})$:
\begin{list}{}{\parsep=0pt\itemsep=7pt\leftmargin=26pt}
\item[(i)]
For any $\chi\in \mbox{Aut}_U(\mbox{CAR})$, $\chi(a_n)$ is expressed in terms 
of a finite polynomial in generators of CAR.
\item[(ii)]
Action of any compact group on CAR is realized by certain elements of 
Aut$_U(\mbox{CAR})$.
\item[(iii)]
Through a nonlinear transformation of CAR, particle numbers are not preserved 
in general.
\item[(iv)]
A subset of Aut$_U(\mbox{CAR})$ consisting of all transformations, which keep 
the Fock vacuum invariant, constitutes a nontrivial subgroup of
Aut$_U(\mbox{CAR})$. 
\item[(v)] 
In general,  $\chi \in \mbox{Au}t_U(\mbox{CAR})$ is an outer automorphism, 
that is, it is {\it not\/} expressed by the adjoint operation of a unitary 
element in CAR as seen by checking the action of $\chi$ on the Fock space.
Restriction of the permutation representation of Cuntz algebra $\calO_{2^p}$
to $\varPhi_{SR_p}(\mbox{CAR})$ determines the change of representations 
of CAR  through the nonlinear transformation.
\end{list}
\vfill\eject
%
%
%

\end{document}